# Global Topological Study of the Protein-protein Interaction Networks

Ka-Lok Ng*

Institute of Bioinformatics

Taichung Healthcare and Management University

No. 500, Lioufeng Road, Wufeng Shiang,Taichung, Taiwan 413

Chien-Hung Huang

Department of Information Management

Ling Tung College

1, Ling Tung Road, Nantun,Taichung, Taiwan 408

## Abstract

We employed the random graph theory approach to analyze the protein-protein interaction database DIP (Feb. 2004), for seven species (*S. cerevisiae, H. pylori, E. coli, C. elegans, H. sapiens, M. musculus and D. melanogaster*). Several global topological parameters (such as node connectivity, average diameter, node connectivity correlation) were used to characterize these protein-protein interaction networks (PINs). The logarithm of the connectivity distribution vs. the logarithm of connectivity study indicated that PINs follow a power law ($P(k) \sim k^{-\gamma}$) behavior. Using the regression analysis method we determined that $\gamma$ lies between 1.5 and 2.4, for the seven species. Correlation analysis provides good evidence supporting the fact that the seven PINs form a scale-free network. The average diameters of the networks and their randomized version are found to have large difference. We also demonstrated that the interaction networks are quite robust when subject to random perturbation. Average node connectivity correlation study supports the earlier results that nodes of low connectivity are correlated, whereas nodes of high connectivity are not directly linked. These results provided some evidence suggesting such correlation relations might be a general feature of the PINs across different species.

# 1. Introduction

Networks of interactions are fundamental to all biological processes; for example, the cell can be described as a complex network of chemicals connected by chemical reactions. Cellular processes are controlled by various types of biochemical networks [1]: (i) metabolic network; (ii) protein-protein interaction network, and (iii) gene regulatory network.

In the last few years, we began to see many progresses in analyzing biological networks using the statistical mechanics of random network approach. The random network approach is becoming a powerful tool for investigating different biological systems, such as the yeast protein interaction network [2], food web [3] and metabolic network [4]. Many studies indicated that there are underlying global structures of those biological networks. Below we highlight the current status of these results.

(I) Metabolic network

Metabolism comprises the network of interaction that provides energy and building blocks for cells and organisms. In many of the chemical reactions in living cells, enzymes act as catalysts in the conversion of certain compounds (substrates) into other compounds (products). Comparative analyses of the metabolic pathways formed by such reactions give important information on their evolution and on pharmacological targets [5]. Recently, the large-scale organization of the metabolic networks of 43 organisms are investigated and it is found that they all have the feature of a scale-free small-world network [6], i.e. $P(k) \sim k^{-\gamma}$, where $k$ is the number of links, and the diameter of the metabolic pathway is the same for the 43 organisms.

(II) Protein-protein interaction network

Proteins perform distinct and well-defined functions, but little is known about how interactions among them are structured at the cellular level. Recently, it was reported that [7] in the yeast organism (a total of 329 proteins by the two-hybrid method measurement [8], the protein-protein interactions are not random, but well organized. It was found that, most of the neighbors of highly connected proteins have few neighbors, that is highly connected proteins are unlikely to interact with each other.

(III) Gene transcription regulatory network

A genetic regulatory network consists of a set of genes and their mutual regulatory interactions. The interactions arise from the fact that genes code for proteins that may control the expression of other genes, for instance, by activating or inhibiting DNA transcription [9]. Recently, it was reported that [10] in the yeast organism, there is a hierarchical and combinatorial organization of transcriptional activity pattern.

# 2. Input data – Database of Interacting Protein (DIP)

There are thousands of different proteins active in a cell at any time. Many proteins act as enzymes, catalyzing the chemical reactions of metabolism. In our analysis we will use the database DIP [11] (http://dip.doe-mbi.ucla.edu) as the input data. DIP is a database that documents experimentally determined protein-protein interactions (a binary relation).

We analyze the DIP database (Feb., 2004), for seven different species, *S. cerevisiae, H. pylori, C. elegans, E. coli, H. sapiens M. musculus* and *D. melanogaster*. In Table 1 we list the statistics of the number of proteins, and number of interactions in our analysis for the seven species.

Table 1. DIP database statistics for *S. cerevisiae, H. pylori, E. coli, C. elegans, H. sapiens, M. musculus and D. melanogaster*.

| Organism | Proteins | Interactions |
|---|---|---|
| *S. cerevisiae (CORE)* | 2631 | 6558 |
| *S. cerevisiae* | 4773 | 15444 |
| *H. pylori* | 710 | 1420 |
| *E. coli* | 429 | 516 |
| *C. elegans* | 2638 | 4030 |
| *H. sapiens* | 946 | 1129 |
| *M. musculus* | 324 | 283 |
| *D. melanogaster* | 7066 | 21017 |

In order to minimize experimental uncertainty, we employed the CORE subset of DIP, which contains the pairs of interacting proteins identified in the budding yeast, *S. cerevisiae* that were validated according to the criteria described in Reference 12.

### 3. Methodology: Topological quantities of a complex network

The biological networks discussed above have complex topology. A complex network can be characterized by certain topological measurements [13]. Erdos and Renyi were the first to propose a model of a complex network known as a random graph [14].

In the graph theory approach, each protein is represented as a node and interaction as an edge. By analyzing the DIP database one constructs an interaction matrix to represent the protein-protein interaction network. In the interaction matrix a value of one and infinity (for convenient a very large number is used in actual programming) is assigned to represent direct interacting and non-interacting protein respectively.

Connectivity distribution and correlation analysis

The first topological feature of a complex network is its degree of connectivity distribution. From the interaction matrix, one can obtain a histogram of *k* interactions for each protein. Dividing each point of the histogram with the number of total number of proteins provide *P(k)*. In a random network, the links are randomly connected and most of the nodes have degrees close to <k>. The degree distribution *P(k)* of a random network with *N* nodes, can be approximated by a Poisson distribution, i.e. $P(k) \approx e^{-<k>} \frac{<k>^K}{k!}$, for large *N* [13, 15]. In many real life networks, the degree distribution has no well-defined peak but has a power-law distribution, $P(k) \sim k^{-\gamma}$, where γ is a constant. Such networks are known as scale-free network. The power-law form of the degree distribution implies that the networks are extremely inhomogeneous. In the scale-free network, there are many nodes with few links and a few nodes with many links. The highly connected proteins could possibly play a key role in the functionality of the network.

In order to quantify the relationship between $P(k)$ and $k^{-\gamma}$, we applied the correlation and regression method to analyze the *Log P(k)* vs *Log k* plot, and calculate the Pearson coefficient $r$, coefficient of determination $r^2$, and the regression coefficient $\gamma$ [16].

Shannon Entropy

To quantitatively characterize the node connectivity distribution, we utilize the Shannon entropy which provide a precise definition of information randomness. Consider a binary sequence X of length *n*, with an element $x_i$ has two states, 0 or 1, the Shannon entropy is given by

$$H(X) = -\sum_{i=1}^{n} p_i \log p_i \qquad (1)$$

where $p_i$, is the probability of observing 0 or 1 with in a given sequence, and *n* is the total degree of freedom (*dof*).

In order to perform a cross-species' node connectivity comparison, we normalized $H(X)$ and consider the relative entropy $H_R$, which is defined as

$$H_R = H / H_{max} \qquad (2)$$

where $H_{max}$ is the Shannon entropy evaluated with uniform probability $p_i$ for all *i*, called. In our PINs study, $p_i$ is equal to $P(k_i)$ and the summation runs over all *dof* with non-zero $P(k_i)$.

Interaction path length, *D*

Proteins can have direct or indirect interaction among themselves [1]. Direct interactions such as binding interactions, including formation of protein complexes, covalent modifications of phosphorylation, glycosylation, and proteolytic processing of polypeptide chains. Indirect interaction refers to two proteins are interacted indirectly via successive chemical reactions. Another class of indirect protein-protein interactions is gene expression, where the message of one protein is transmitted to another protein via the process of protein synthesis from the gene.

The second topological measurement is the distance between two nodes, which is given by the number of links along the shortest path. The number of links by which a node is connected to the other nodes varies from node to node. The diameter of the network, also known as the average path length, is the average of the distances between all pairs of nodes.

For all pairs of proteins, the shortest interaction path length, *j* (i.e. the smallest number of reactions by which one can reach protein 2 from protein 1) will be determined by using the Floyd's algorithm [17]. Floyd's algorithm is an algorithm to find the shortest paths for each vertex in a graph. The algorithm represents a network having *N* nodes as an *N* by *N* square matrix *M*. Each entry (*i,j*) of the matrix gives the distance $d_{ij}$ from node *i* to node $j$. If $i$ is linked directly to $j$, then $d_{ij}$ is finite, otherwise, it is infinite. Floyd's algorithm is based upon the idea that, given three nodes $i$, $j$ and $k$, it is shorter to reach $j$ from $i$ by passing through $k$ if $d_{ik} + d_{kj} < d_{ij}$.

The average diameter $d$ of a network is given by

$$d = \frac{\sum_j j f(j)}{\sum_j f(j)} \qquad (3)$$

where $j$ is the shortest path length and $f(j)$ is the frequency of nodes have path length $j$.

In order to compare the interaction network with the randomized version of the original network; we arbitrary assigned an interaction between any two proteins (nodes) in our simulation, and calculate the multiple sampling (160 times of random sampling) average diameter $<d_{rand}>$, while keeping the number of nodes and interactions (edges) same as the original network.

Robustness of network

In order to test whether the interaction network is robust against errors, we followed the method introduced in Reference 7, and slightly perturbed the network randomly. First, we randomly select a pair of edges A-B and C-D. The two edges are then rewired in such a way that A connected to D, while C connected to B. Notice that this process will not change the degree of connectivity of each node. A repeated application of the above step leads to a randomized version of the original network. Multiple sampling (160 times of random sampling) of the randomized networks allowed us to calculate the average diameter of the perturbed network $<d_{pert}>$ and compare the perturbed result with the unperturbed network diameter, i.e. $\Delta = (<d_{pert}> - d)/d*100\%$.

Correlation profile of node connectivity

In the random graph models with arbitrary degree distribution the node degrees are uncorrelated [15]. In order to test for the node connectivity correlations, we employed the approach introduced in Reference 7, and count the frequency $P(K_1,K_2)$ that two proteins with connectivity $K_1$ and $K_2$ connected to each other by a link and compared it to the same quantity $P_R(K_1,K_2)$ measured in a randomized version of the same network. The average node connectivity $\langle K_2 \rangle$ for a fixed $K_1$ is given by,

$$\langle K_2 \rangle = \sum K_2 \frac{P(K_1,K_2)}{\langle P_R(K_1,K_2) \rangle} \qquad (4)$$

where $\langle P_R(K_1,K_2) \rangle$ is the multiple sampling average (160 times of random sampling), and the summation sums for all $K_2$ with a fixed $K_1$. In the randomized version, the node connectivity of each protein are kept the same as in the original network, whereas their linking partner is totally random.

## 4. Results

In Table 2, we present the average connectivity $<k>$, maximum connectivity $k_{max}$, average diameter $d$, average random diameter $<d_{rand}>$, average perturbed diameter $<d_{pert}>$ and the $\Delta$ results for the seven species.

Table 2. The average connectivity $<k>$, maximum connectivity $k_{max}$, average diameter $d$, average random diameter $<d_{rand}>$, average perturbed diameter $<d_{pert}>$ and the $\Delta$ results for the seven species.

| Species | $<k>$ | $k_{max}$ | $d$ | $<d_{rand}>$ | $<d_{pert}>$ ($\Delta$) |
|---|---|---|---|---|---|
| *S. cerevisiae(CORE)* | 4.87 | 111 | 5.01 | 4.01 | 5.01 (0.0%) |
| *S. cerevisiae* | | 283 | 4.19 | 3.67 | 4.19 (0.0%) |
| *H. pylori* | 3.87 | 54 | 4.14 | 3.72 | 4.14 (0.0%) |
| *E. coli* | 3.02 | 54 | 3.22 | 5.94 | 3.40 (5.6%) |
| *C.elegans* | 1.91 | 187 | 4.81 | 4.43 | 4.81(0.0%) |
| *H. sapiens* | 2.30 | 33 | 6.05 | 5.61 | 6.16 (1.8%) |
| *M. musculus* | 1.67 | 12 | 3.58 | 6.64 | 3.74 (4.7%) |
| *D. melanogaster* | 5.90 | 80 | 4.46 | 5.20 | 4.46(0.0%) |

The average diameter $d$ of the networks lies between 3.2 to 6.0 for any two proteins. We noticed there is a large difference between $d$ and $<d_{rand}>$ for each species, hence we concluded that the protein-protein interaction networks are mostly likely not random. One can also conclude from Table 2 that the interaction network is quite robust when subject to random perturbation, that is $<d_{pert}>$ is slightly differs from $d$, that is a small $\Delta$ value. This can be interpreted that the protein-protein interaction network is robust against external perturbation [7].

We plot the logarithm of the normalized connectivity distribution $Log(P(k))$ as a function of the logarithm of connectivity $Log\ k$, for single cellular organisms (*E. coli, H. pylori* and *S. cerevisiae(CORE)*) and multi-cellular organisms (*C. elegans, D. melanogaster, H. sapiens* and *M. musculus*) in Figure 1 and Figure 2 respectively.

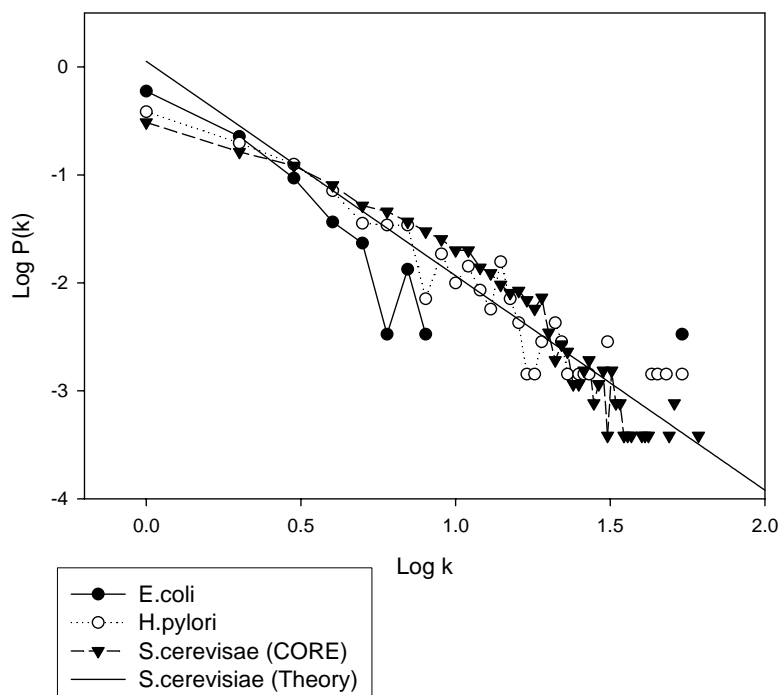

Fig.1. The logarithm of normalized frequency of connectivity vs the logarithm of connectivity for single cellular organisms, *E. coli, H. pylori*, *S. cerevisiae(CORE)* and *S. cerevisiae(CORE, Theory)*.

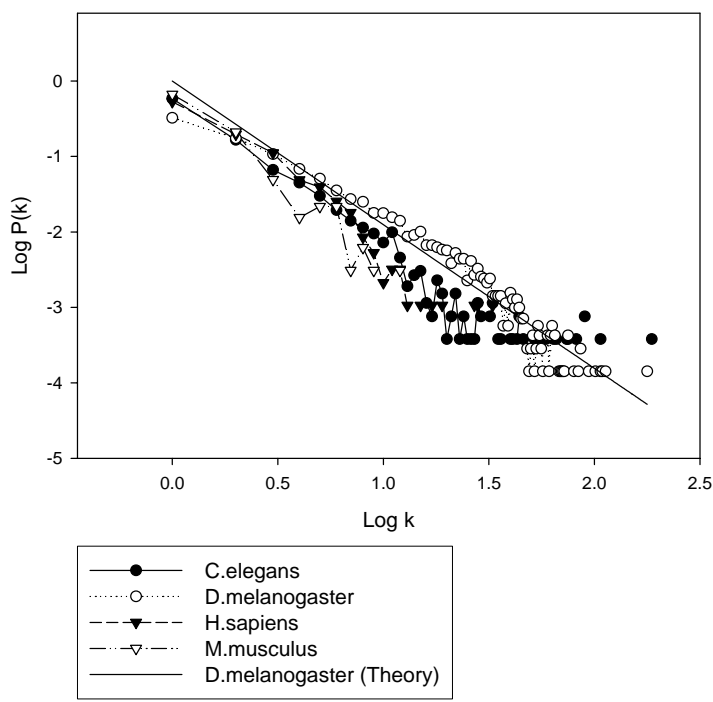

Fig. 2 The logarithm of normalized frequency of connectivity vs the logarithm of connectivity for multiple cellular organisms, *C. elegans, D. melanogaster, H. sapiens, M. musculus* and *C. elegans(Theory)*.

It is evident from the figures that the number of proteins decrease with increasing number of connections, that is it has an inverse relation. In other words, protein has multiple connections are rare. The logarithm of the connectivity distribution vs. the logarithm of connectivity indicated that PINs follow a power law ($P(k) \sim k^{-\gamma}$) behavior. It suggests that the PINs form a scale-free network. The scale-free degree distribution conclusion is reported in certain pervious publications [7,18].

In Table 3, we present the regression and correlation analysis for the node connectivity distribution data. Using the regression analysis method, we determined that $\gamma$ lies between 1.5 and 2.4, for the seven species we considered. For yeast, the regression coefficient $\gamma$ we obtained is consistent with the result reported in Reference 7 ($\gamma =2.5\pm0.3$), where the authors considered nuclear proteins only. Correlation analysis provides good evidence supporting the fact that the seven PINs form a scale-free network. The closer $r^2$ is to 1, the better it accounts for the correlation [16]. Relative to other species, *E. coli* has a smaller $r^2$ (i.e. 0.70), therefore, the correlation is not as strong as the other species.

Table 3. The regression coefficient $\gamma$, Pearson coefficient $r$, coefficient of determination $r^2$, total degree of freedom $n$, and the relative Shannon entropy $H_R$ for the DIP data.

| Species | $\gamma$ | $r$ | $r^2$ | Total dof, $n$ | $H_R$ |
|---|---|---|---|---|---|
| *S. cerevisiae(CORE)* | 2.0±0.1 | 0.95 | 0.91 | 44 | 0.601(0.635) |
| *H. pylori* | 1.7±0.1 | 0.95 | 0.90 | 30 | 0.606 |
| *E. coli* | 1.5±0.4 | 0.84 | 0.70 | 9 | 0.541 |
| *C. elegans* | 1.6±0.1 | 0.92 | 0.84 | 49 | 0.405 |
| *H. sapiens* | 2.1±0.1 | 0.96 | 0.93 | 19 | 0.513 |
| *M. musculus* | 2.4±0.2 | 0.97 | 0.93 | 10 | 0.454 |
| *D. melanogaster* | 1.90±0.02 | 0.96 | 0.93 | 76 | 0.577 |

The relative Shannon entropy result for the seven species is reported in Table 3 and depicted in Figure 3 as well. The relative Shannon entropy is 0.601 and 0.635 for *S. cerevisiae* and the CORE subset of *S. cerevisiae* respectively. Shannon entropy calculation of the connectivity distribution seems to suggest that multi-cellular organisms (except *D. melanogaster*) tend to have a lower entropy value compare to single cellular organisms. It is known that Shannon entropy is a measure of uncertainty, the lower the entropy (uncertainty), the more structure is embedded in the data.

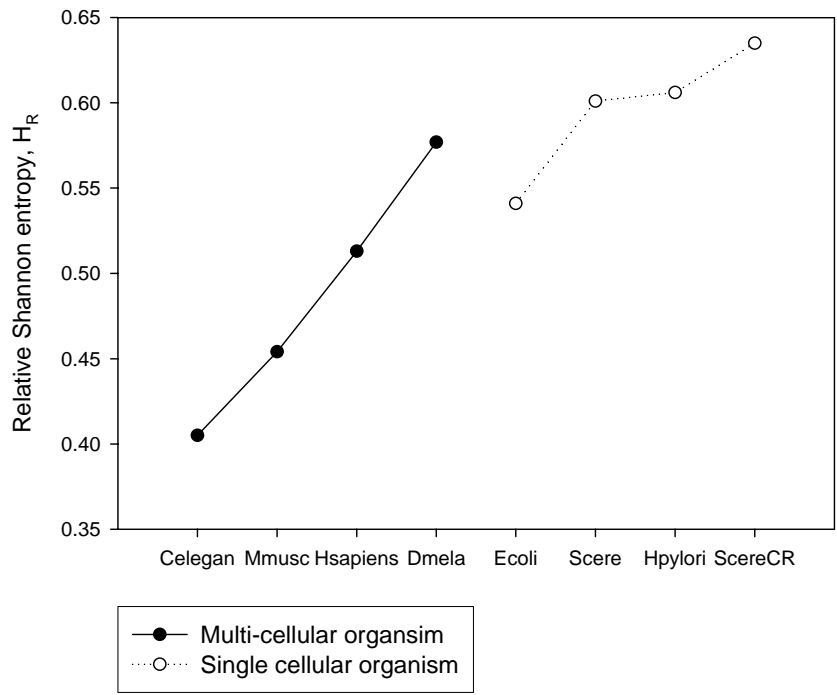

Fig. 3. The relative Shannon entropy for single cellular organisms (*S. cerevisiae, S. cerevisiae (CORE), H. pylori* and *E. coli*) and the multi-cellular organisms *(C. elegans, H. sapiens, M. musculus* and *D. melanogaster*).

In Fig. 4, we plot the logarithm of the normalized frequency of connected paths vs the logarithm of their length.

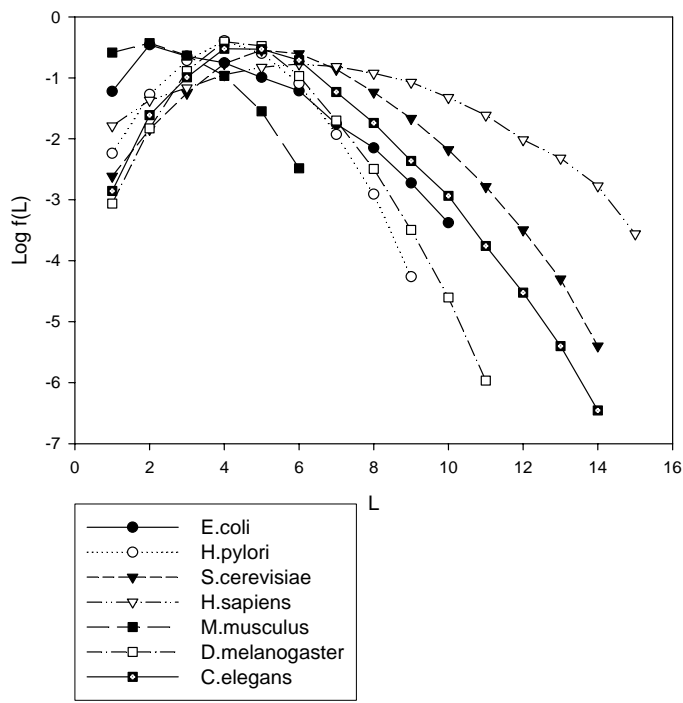

Fig. 4. The logarithm of the normalized frequency distribution of connected paths vs the logarithm of their length for *S. cerevisiae(CORE), H. pylori, E. coil, H. sapiens, M. musculus and D. melanogaster*.

In Table 4 we present the longest path length $L_{max}$, the total frequency of $L_{max}$, $f_{max}$, and the length of the path with the highest frequency $L'$ results, for the seven species. It is found that *H. sapiens* tend to have a higher $L_{max}$ and $L'$ values. A larger $L_{max}$ and $L'$ values mean that any two proteins can have indirect interaction via more successive chemical reactions. We also noticed that one could approximate network diameter $d$ by the network's highest frequency path length $L'$, that is, the $L'$ values in Table 4 is approximately equal to the $d$ values in Table 2.

Table 4. The longest path length $L_{max}$, their total frequency of occurrence $f_{max}$, and path length with the highest frequency $L'$ results, for *S. cerevisiae(CORE), H. pylori, E. coil, C. elegans, H. sapiens, M. musculus* and *D. melanogaster*.

| Species | $L_{max}$ | $f_{max}$ | $L'$ (by percentage) |
|---|---|---|---|
| *S. cerevisiae(CORE)* | 13 | 8 | 5 (33%) |
| *S. cerevisiae* | 12 | 2 | 4 (45%) |
| *H. pylori* | 9 | 26 | 4 (40%) |
| *E. coli* | 10 | 4 | 2 (35%) |
| *C.elegans* | 14 | 2 | 4(30%) |
| *H. sapiens* | 16 | 8 | 5 (17%) |
| *M. musculus* | 11 | 52 | 4 (39%) |
| *D. melanogaster* | 14 | 40 | 4(19%) |

From Figure 5 to 6, we showed the logarithm of the average degree connectivity $Log(<K_2>)$ vs the node connectivity $K_1$ for the single and multi-cellular species. These results indicated that there is a strong correlation between nodes of lower connectivity; that is the upper left region of every plot. In the large $K_1$ or lower half region of every plot, there are a few scattered points, which suggest that nodes of high connectivity are link to nodes of lower connectivity, whereas nodes of high connectivity are not directly linked. In fact, the $<K_2>$ value evaluated at $K_1=1$ is of the order of 10 to 30 folds larger than $<K_2>$ at $K_1 = k_{max}$ for the seven species. The absence of any structures in the large $K_1$ region of the figures strongly indicated that nodes of high connectivity are not directly linked. The same conclusion is obtained previously in Reference 7 for the species yeast only. Our analysis extended the previous work [7] to seven species, hence, these results provided some evidence suggesting such correlation relations might be a general feature of the PINs across different species.

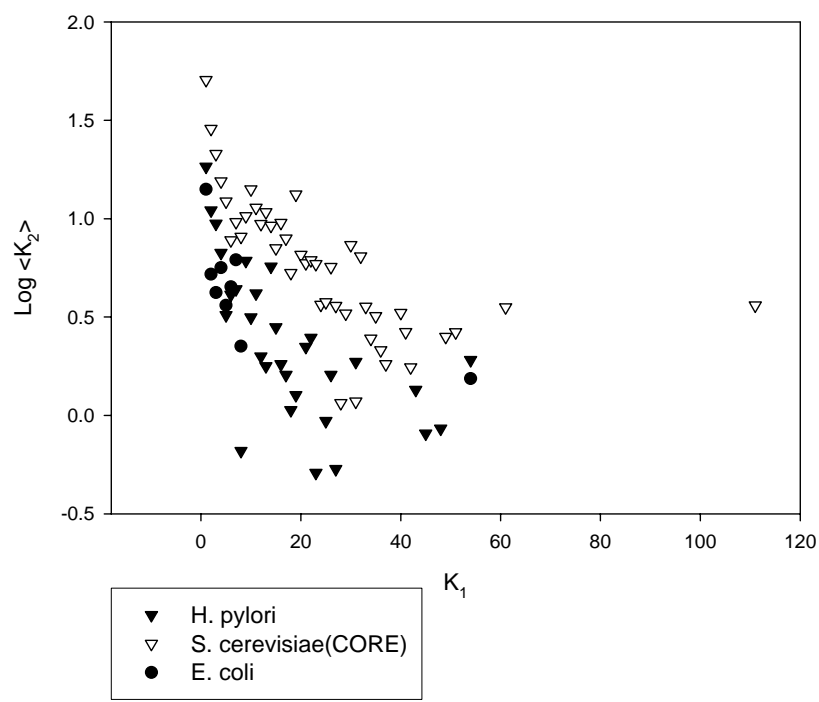

Fig. 5. Average node degree correlation profile of protein-protein interaction network for the single cellular organisms, *H. pylori, S. cerevisiae(CORE)* and *E. coli*.

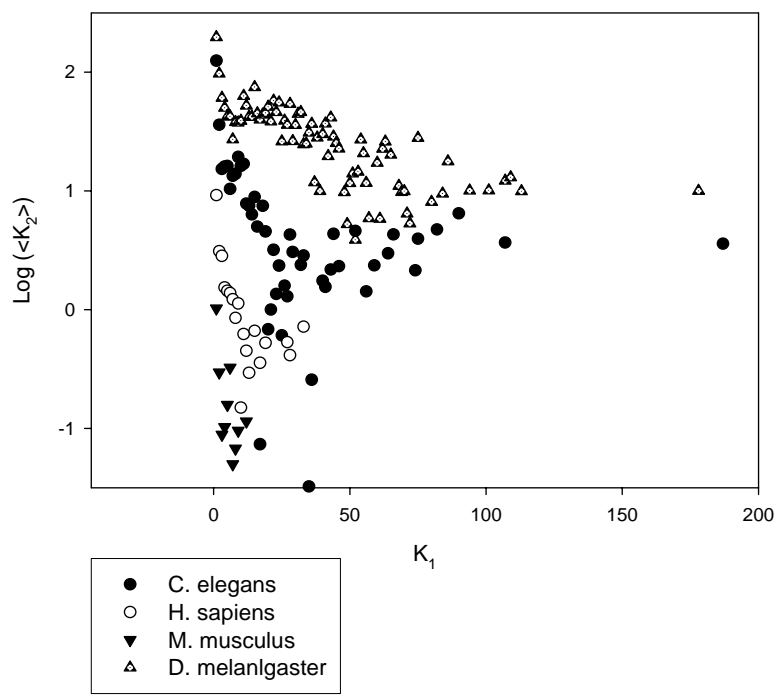

Fig. 6. Average node degree correlation profile of protein-protein interaction network for multiple cellular organisms, *C. elegans, D. melagonaster, H. sapiens* and *M. musculus*.

# 5. Conclusions and Discussions

We employed the random graph theory approach to analyze the latest version of the protein-protein interacting database DIP (Feb. 2004), for seven different species (*S. cerevisiae, H. pylori, E. coli, C. elegans, H. sapiens, M. musculus* and *D. melanogaster*).

Several global topological parameters (such as node connectivity, average diameter, node connectivity correlation) were used to characterize these protein-protein interaction networks (PINs). The logarithm of the connectivity distribution *Log(P(k))* vs. the logarithm of connectivity *Log k* indicated that PINs follow a power law ($P(k) \sim k^{-\gamma}$) behavior. Using the regression analysis method, we determined that γ lies between 1.5 and 2.4, for the seven species. Correlation analysis provides good evidence supporting the fact that the seven PINs form a scale-free network. We also adapt the Shannon entropy approach to quantify the connectivity distribution, the result seems to suggest that multi-cellular organisms tend to have a lower relative entropy value for the single cellular organisms.

In order to compare the interaction networks with the random network, a randomized version of the original interaction network is generated using Monte Carlo simulation. The average diameters *d* of the networks and their randomized version $<d_{rand}>$ are found to have large difference, hence we concluded that PINs are mostly likely not random. We also demonstrated that the interaction networks are quite robust when subject to random perturbation, that is the average diameter for the perturbed case, $<d_{pert}>$ are slightly differ from the unperturbed cases, *d*. All the interacting paths, either direct or indirect are identified for the seven species.

Average node connectivity correlation study supports the earlier results that nodes of low connectivity are correlated, whereas nodes of high connectivity are not directly linked. Our analysis extended the previous work [7] to seven species, hence, these results provided some evidence suggesting such correlation relations might be a general feature of the PINs across different species.

Here we focused on two-body interactions and it will be interesting to consider multi-body interactions in the protein network and find clusters of proteins that have many interactions among themselves [19]. Such clusters correspond to protein complexes. Another interesting area of work is to show that if two proteins share significantly large number of common partners than random, they could have close functional associations [20]. Furthermore, the PINs provide a binary interaction information only, it dose not necessary means that within the cell, the interacted proteins will be presented in the same localization and time, therefore, it would be more helpful if one takes the spatial (subcellular localization) and temporal effects into consideration.

# Acknowledgements

One of the authors, Ka-Lok Ng, would like to thank the Institute of Physics, Academia Sinica to support this work through a summer visiting funding. This work is supported in part by the R.O.C. NSC grant NSC 92-2112-M-468-001.